\documentclass[aps,prd,twocolumn,nofootinbib,superscriptaddress]{revtex4-2}

\pdfoutput=1

\usepackage[utf8]{inputenc} 
\usepackage{amsmath}
\usepackage{amssymb}
\usepackage{xcolor}
\usepackage{graphicx}
\usepackage{dcolumn}
\usepackage{bm}
\usepackage{microtype}
\usepackage{threeparttable}
\usepackage{mathtools}
\usepackage[colorlinks=true]{hyperref}
\usepackage{physics}

\newcommand{\squeezeup}{\vspace{-4mm}}

\begin{document}

\title{Multitrace deformations and the nonlinear stability of Anti-de Sitter space}

\author{Alexandre Serantes}
\email{alexandre.serantesrubianes@ugent.be}
\affiliation{Department of Physics and Astronomy, Ghent University, 9000 Ghent, Belgium}

\author{David Travieso Mayo}
\email{david.travieso.mayo@usc.es}
\affiliation{Departamento de Física de Partículas, Universidade de Santiago de Compostela and Instituto Galego de Física de Altas Enerxías (IGFAE), E-15782 Santiago de Compostela, Spain}

\author{Javier Mas}
\email{javier.mas@usc.es}
\affiliation{Departamento de Física de Partículas, Universidade de Santiago de Compostela and Instituto Galego de Física de Altas Enerxías (IGFAE), E-15782 Santiago de Compostela, Spain}

\begin{abstract}

\noindent We investigate the nonlinear stability of global Anti-de Sitter space in the presence of multitrace deformations utilizing an Einstein-Klein-Gordon system with a top-down scalar potential. Our numerical simulations show that marginal and irrelevant deformations retain the nonlinear instability originally found by Bizo\'n and Rostworowski, while relevant deformations disrupt it. We show that the nonlinear impact of multitrace deformations is in one-to-one correspondence with how they affect the resonant character of the global Anti-de Sitter normal mode eigenfrequency spectrum.

\end{abstract}

\maketitle

\section{Introduction}
\label{introduction}

Among the three maximally symmetric spacetimes, Minkowski and de Sitter are known to be nonlinearly stable. The nonlinear stability of global Anti-de Sitter (AdS) space has been under intense scrutiny since the seminal work of P.~Bizo\'n and A.~Rostworowski  in 2011 \cite{Bizon:2011gg}. %Cite nonlinear stability of Minkowski and de Sitter
Ref.~\cite{Bizon:2011gg} considered spherically symmetric, asymptotically global AdS$_4$  geometries supported by a massless scalar field and found that initial data consisting of a Gaussian with small width $\sigma$ always underwent gravitational collapse, no matter how small their amplitude $\varepsilon$ was. The mechanism behind this nonlinear instability was a turbulent cascade of energy to progressively smaller spatial scales, which eventually lead to the formation of a black hole on times scaling as $\varepsilon^{-2}$ in the $\varepsilon \to 0$ limit. We will refer to this phenomenon as the \textit{Bizo\'n-Rostworowski (BR) instability}.

One of the key lessons that has been learned in subsequent studies of the nonlinear stability of global AdS is that, for an BR instability to be present, there has to be the possibility of understanding the initial data as a small perturbation of a geometry with an exactly resonant normal mode eigenfrequency spectrum \cite{Dias:2012tq}, such that the turbulent cascade can happen. This condition is satisfied in the AdS vacuum, but fails in time-periodic geometries such as oscillons \cite{Maliborski2013, Green:2015dsa} and boson stars \cite{Buchel:2013uba, Maliborski:2014rma}, or static geometries such as solitons in Einstein-Maxwell-scalar theory \cite{Arias:2016aig}. The basins of attraction of these non-vacuum geometries act as \emph{islands of nonlinear stability} in the space of initial data: for any initial data within these islands, the turbulent cascade is not at work and gravitational collapse is averted.

The approach to studying the role of the resonant character of the eigenfrequency spectrum in the BR instability described in the previous paragraph can be though of as proceeding in three steps. First, one identifies a special excited state exhibiting either discrete or continuous time-translational invariance, such that concept of a normal mode is applicable. Second, one checks by an explicit computation that the resonant property of the AdS vacuum spectrum is spontaneously broken in this excited state. Finally, it remains to assess whether the nonlinearly stable initial data of interest can indeed be understood as a special excited state subject to a finite, but sufficiently small, perturbation.

In this work, we propose an alternative route. Rather than focusing on stability islands sourced by special excited states in our original theory, we will instead consider a deformed theory where the boundary conditions at asymptotic infinity are modified and the resonant character of the AdS vacuum spectrum is explicitly broken from the very beginning. This allows studying the interplay of the resonance condition and the BR instability in initial data corresponding to a \emph{bona fide} perturbation of the AdS vacuum. 

To implement this strategy, we will work with a top-down AdS$_{4}$-Einstein-scalar system such that the scalar field mass is in the window
\begin{equation}
m_{BF}^2 \leq m^2 < m_{BF}^2 + \frac{1}{L^2},\quad m_{BF}^2 = - \frac{9}{4 L^2},      
\end{equation}
and impose designer gravity boundary conditions.\footnote{See Refs.~\cite{Masachs:2019znp,Bizon:2020yqs,Ficek:2023qni} for recent studies of AdS-Einstein-scalar systems with a particular kind (Robin) of designer gravity boundary conditions.} This approach can be thought of as the natural analog of the flat space problem discussed in Refs.~\cite{Maliborski:2012gx,Maliborski:2014rma,Kurzweil:2022red}, where a spherically symmetric geometry supported by a massless scalar field enclosed in a box was considered. In this flat space case, the choice between Dirichlet or Neumann boundary conditions for the scalar field is in one-to-one correspondence with the resonant or non-resonant nature of the spectrum and the presence or absence of the BR instability. 

In the context of holography, designer gravity boundary conditions correspond to multitrace deformations of the boundary CFT \cite{Witten:2001ua, Papadimitriou:2007sj}. In the past, multitrace deformations have been employed to engineer unbounded effective field theory potentials supporting big crunch cosmologies in the bulk \cite{Hertog:2004rz, Hertog:2005hu}, quantum quenches \cite{Basu:2013soa} or phase transitions \cite{Ares:2021ntv, Ares:2021nap, Caddeo:2024lfk}. Our results on the nonlinear stability of AdS$_4$ with designer gravity boundary conditions can be regarded as exploring the effect of multitrace deformations on the thermalization of spatially homogeneous states in ABJM theory \cite{Aharony:2008ug} placed on a two-sphere.

The outline of this paper is as follows. In Sec.~\ref{sec:model}, we introduce the AdS-Einstein-scalar system we work with. Then, in Sec.~\ref{sec:linear_stability}, we explain how multitrace deformations affect the eigenfrequency spectrum, to finally present our novel results on the nonlinear stability of global AdS$_4$ in Sec.~\ref{sec:nonlinear_stability}. We conclude in Sec.~\ref{sec:discussion} with a discussion. We have relegated various technical aspects of our work to two appendices. App.~\ref{app:eom} contains the equations of motion of our AdS-Einstein-scalar system and explains how to put them in a form suitable for numerical implementation, while App.~\ref{app:numerics} provides a comprehensive discussion of our simulation code, including convergence tests. 

\section{Multitrace deformations in an Einstein-scalar system}
\label{sec:model}

We consider a holographic CFT whose bulk dual is given by Einstein gravity with negative cosmological constant and a massive scalar field $\phi$, 
\begin{equation}
\begin{split}\label{action}
S = &\frac{1}{16 \pi G} \int d^4 x \sqrt{-g} \left(R + 6 - \frac{1}{2}(\nabla\phi)^2 - V(\phi)\right) + \\
&S_{GHY}+ S_{ct},
\end{split}
\end{equation}
with potential  
\begin{equation}\label{potential}
V(\phi) = -4 \sinh\left(\frac{\phi}{2}\right)^2.    
\end{equation}
$S_{GHY}$ and $S_{ct}$ are respectively the Gibbons-Hawking-York and counterterm actions, and we have set the AdS radius $L=1$. The action \eqref{action} arises as a consistent truncation of eleven-dimensional supergravity on AdS$_4 \times S^7/Z_k$, and has a dual interpretation as a particular sector of ABJM theory \cite{Cvetic:1999xp, Costa:2017tug}. %Check this!

The potential \eqref{potential} is such that the scalar field $\phi$ has a negative mass-squared $m^2 =-2$ and belongs to the double-quantization window:  near the AdS$_4$ boundary, located at $r \to \infty$ in Schwarzschild coordinates, $\phi$ falls off as 
\begin{equation}
\phi = \frac{\alpha}{r} + \frac{\beta}{r^2} + \ldots
\label{falloff}    
\end{equation}
and both the boundary conditions  
\begin{equation}
\alpha = 0,~\beta~\textrm{free (standard quantization)}
\label{stand_quant}
\end{equation}
and 
\begin{equation}
\alpha~\textrm{free},~\beta=0~\textrm{(alternative quantization)}
\end{equation}
lead to normalizable modes and correspond to a well-defined dual single-trace primary scalar operator $\mathcal{O}$. In the standard quantization, $\mathcal O$ has conformal dimension $\Delta = 2$ and $\langle \mathcal{O} \rangle \propto \beta$; in the alternative quantization, $\mathcal O$ has conformal dimension $\Delta = 1$ and $\langle \mathcal{O} \rangle \propto \alpha$. 

Our focus, however, is not in the standard quantization nor in the alternative one. It turns out that the scalar field allows for \textit{designer gravity} boundary conditions \cite{Hertog:2004dr,Hertog:2004ns} of the form 
\begin{equation}\label{bc:general}
\beta = \beta_{BC}(\alpha),     
\end{equation}
where $\beta_{BC}$ is an arbitrary function. The CFT interpretation of the designer gravity boundary conditions is as follows: defining a function ${\cal W}(\alpha)$ such that 
\begin{equation}
\beta_{BC}(\alpha) = \frac{\delta \mathcal{W}(\alpha)}{\delta\alpha},     
\end{equation}
the designer gravity boundary conditions \eqref{bc:general} correspond to the following modification of the dual CFT Lagrangian, 
\begin{equation}\label{multitrace_def}
\mathcal{L}_{CFT} \to \mathcal{L}_{CFT} + \mathcal{W}(\mathcal{O})\, .   
\end{equation}
Here $\mathcal{O}$ is understood as the operator dual to $\phi$ in the alternative quantization. In general,
\begin{equation}
\mathcal{W}(\mathcal{O}) = \sum_{n=1}^\infty \frac{c_n}{n} \mathcal{O}^n.      
\end{equation}
Given that $\mathcal{O}$ is a single-trace primary operator, the expression above justifies referring to Eq.\,\eqref{multitrace_def} as a \textit{multitrace deformation} of the original CFT. Note also that, since $\Delta = 1$: 
\begin{itemize}
\item $\mathcal{O}$, $\mathcal{O}^2$ are relevant deformations. 
\item $\mathcal{O}^3$ is a marginal
deformation. This triple-trace deformation corresponds to a designer gravity boundary condition of the form 
\begin{equation}
\beta_{BC}(\alpha) = c_3\alpha^2,       
\end{equation}
and is singled out by the property that it does not destroy the algebra of asymptotic isometries of global AdS$_4$ (see Refs. \cite{Hertog:2004ns, Anabalon:2015xvl}). 
\item $\mathcal{O}^n$ with $n \geq 4$ are irrelevant deformations. 
\end{itemize}
In this work, we specialize to spherically symmetric geometries of the form  
\begin{subequations}\label{time_ansatz}
\begin{equation}
\hspace{-0.5em}ds^2 = \frac{\left(-f(t,x) e^{-2\delta(t,x)}dt^2 + \frac{dx^2}{f(t,x)} + \sin^2 x d\Omega_2^2\right)}{\cos^2 x}, 
\end{equation}
\begin{equation}
\phi = \phi(t,x), 
\end{equation}
\end{subequations}
where $x \in [0, \pi/2)$ parameterizes the radial direction, with $x=0$ being the origin and $x=\frac{\pi}{2}$ the asymptotic boundary. The equations of motion stemming from the ansatz \eqref{time_ansatz} are presented in Appendix \ref{app:eom}. According to them, the geometry \eqref{time_ansatz} is fully determined by values of the scalar field $\phi(t,x)$ and the conjugate momentum,  
\begin{equation}
\Pi(t,x) = \frac{e^{\delta(t,x)}}{f(t,x)}\partial_t \phi(t,x),  
\end{equation}
at $t=0$. Global AdS$_4$  corresponds to a vanishing $\phi(t,x) = 0$ and is given by $f(t,x) = 1$ and $\delta(t,x) = 0$. 

\section{Linear stability}
\label{sec:linear_stability}

Before discussing the nonlinear stability of global AdS$_4$ with designer gravity boundary conditions, we comment on the normal mode spectrum (see also Ref.~\cite{Bizon:2020yqs}). To compute the normal modes, we put forward the following power series ansatz for the geometry, 
\begin{subequations}\label{power-series_ansatz}
\begin{gather}
\phi(t,x) = \sum_{n=1}^\infty \phi_n(t,x) \gamma^n, \\
f(t,x)= 1 + \sum_{n=1}^\infty f_n(t,x) \gamma^n, \\
\delta(t,x)=\sum_{n=1}^\infty \delta_n(t,x) \gamma^n, 
\end{gather}
\end{subequations}
and solve the equations of motion at leading order in $\gamma \to 0$. Assuming that $c_n = O(1)$ and $\alpha = O(\gamma)$ in the $\gamma \to 0$ limit, it follows that $\beta = o(\gamma)$ for marginal and irrelevant deformations, while $\beta = O(\gamma)$ for relevant double-trace deformations. 

The equation of motion for $\phi_1$ is  
\begin{equation}
-\ddot{\phi}_1 + \phi_1'' + 2 \sec x \csc x \phi_1' + 2 \sec^2 x \phi_1 = 0, 
\end{equation}
where dots (primes) denote temporal (spatial) derivatives. This equation allows for plane-wave solutions of the form
\begin{equation}\label{plane-wave_solution}
\phi_1(t,x) = \frac{A \sin(\omega x) + B \cos(\omega x)}{\tan x} \cos \omega t.
\end{equation}
Regularity of $\phi(t,x)$ at $x=0$ requires that $B=0$. On the other hand, given Eq.~\eqref{plane-wave_solution}, the series expansion of $\phi$ near the asymptotic boundary up to $O(\gamma)$ reads ($y \equiv \frac{\pi}{2}-x$)
\begin{equation}\label{bound_A}
\gamma \left[A \sin\left(\frac{\pi}{2} \omega\right) y{-} A\, \omega \cos\left(\frac{\pi}{2} \omega\right) y^2 + O(y^3) \right],
\end{equation}
and has to match the general expression \eqref{falloff}, 
\begin{equation}\label{bound_B}
\alpha y + \beta y^2 + \mathcal{O}(y^3), 
\end{equation}
resulting in 
\begin{subequations}
\begin{gather}
\alpha = \gamma A \sin\left(\frac{\pi}{2} \omega\right) + O(\gamma^2),  \\
\beta = - \gamma A\, \omega \cos\left(\frac{\pi}{2} \omega\right) + O(\gamma^2). 
\end{gather}
\end{subequations}
For non-relevant deformations, $\beta = o(\gamma)$, and hence the normal mode eigenfrequencies $\{\omega_n, n \in \mathbb N\}$ are the roots of $\cos\left(\frac{\pi}{2} \omega\right) = 0$, 
\begin{equation}\label{spectrum_alt}
\omega_n = 2 n - 1,~n \in \mathbb N 
\end{equation}
This is the same eigenfrequency spectrum as in the alternative quantization, and it is exactly resonant. In contrast, for relevant double-trace deformations, $\beta=c_2 \alpha$, the normal mode eigenfrequencies are the roots of a transcendental equation 
\begin{equation}\label{eq:eigenfr}
\frac{c_2 \pi}{2} \mathrm{sinc} \left(\frac{\pi}{2} \omega\right) + \cos\left(\frac{\pi}{2} \omega\right) = 0.  
\end{equation}
In the absence of the double-trace deformation, $c_2=0$, we recover the normal mode eigenfrequency spectrum of the alternative quantization, Eq.~\eqref{spectrum_alt}. In the opposite, $c_2 \to \infty$ limit, we recover the normal mode eigenfrequency spectrum of the standard quantization, 
\begin{equation}
\omega_n = 2n,~n \in \mathbb N.      
\end{equation}
Analytical control over the full eigenfrequency spectrum can be gained by solving Eq.~\eqref{eq:eigenfr} in a perturbative expansion in $c_2 \to 0$. We find 
\begin{equation}
\omega_n = \omega_n^{(0)} + \frac{2}{\pi \omega_n^{(0)}} c_2 - \frac{4}{\pi^2 \omega_n^{(0)^3}} c_2^2 + O(c_2^3),   
\end{equation}
where $\omega_n^{(0)} = 2n-1,~n \in \mathbb N$ is the alternative quantization eigenfrequency spectrum. Assuming that the $c_2 \to 0$ and $n \to \infty$ limits commute, the expression above results in 
\begin{equation}
\omega_{n+1} - \omega_n = 2 - \frac{c_2}{\pi n^2} + O\left(\frac{1}{n^4}\right)    
\end{equation}
and implies that the eigenfrequency spectrum is not exactly resonant, but only asymptotically so. This assumption can be verified by an explicit numerical computation. Note that both for oscillons and boson stars in global AdS$_4$ the difference between two consecutive eigenfrequencies is $O\left(\frac{1}{n}\right)$ at large $n$ \cite{Maliborski:2014rma,Green:2015dsa}, while here it is $O\left(\frac{1}{n^2}\right)$. %Check this again
%\squeezeup
\begin{figure}[h!]
\centering
\includegraphics[width=\linewidth]{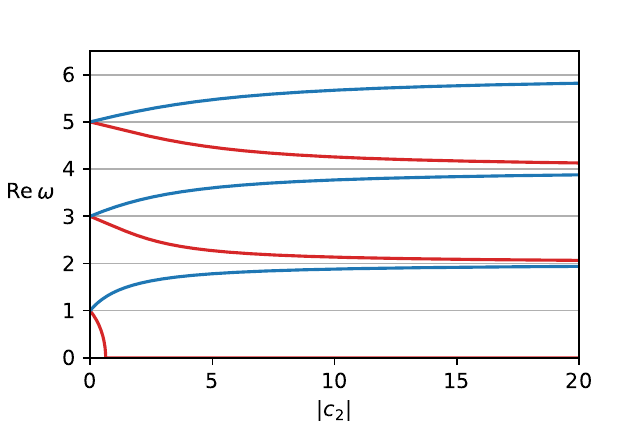}
\caption{\small Eigenfrequencies for double-trace deformations as function of $c_2$. Blue (red) curves correspond to positive (negative) values of $c_2$. For $c_2 \to 0$ ($c_2\to\infty$), we recover the alternative (standard) quantization frequencies. At $c_2=-2/\pi$, the first eigenfrequency vanishes. For lower values of $c_2$, this eigenfrequency is purely imaginary, signaling a linear instability.}
\label{fig:normal_modes}
\end{figure}

Finally, we note that in the $c_2 \to - \frac{2}{\pi}$ limit the system supports a zero mode, as can be immediately seen by expanding Eq.~\eqref{eq:eigenfr} around $\omega = 0$. The existence of the zero mode allows for an analytical description of the lowest-lying eigenfrequency in the vicinity of $c_2 = - \frac{2}{\pi}$, 
\begin{equation}\label{unstable_mode}
\begin{split}
\omega_1 = &\left(\frac{6}{\pi} \right)^\frac{1}{2}\left(c_2 + \frac{2}{\pi} \right)^\frac{1}{2}\left[1 - \frac{\pi}{20} \left(c_2 + \frac{2}{\pi} \right)\right. + \\
& \left.\frac{9 \pi^2 }{5600}\left(c_2 + \frac{2}{\pi} \right)^2 + \ldots \right].
\end{split}
\end{equation}
Note that, for $c_2 < -\frac{2}{\pi}$, $\omega_1 = i \lambda$ with $\lambda\in \mathbb R$ and the system becomes linearly unstable. 

Away from the limits $c_2 \to 0$, $c_2 \to \infty$ and $n \to \infty$, the full eigenfrequency spectrum has to be determined numerically. The reader can find the first three eigenfrequencies as a function of $c_2$ in Fig.~\ref{fig:normal_modes}, where the linear instability kicking in at $c_2 = -\frac{2}{\pi}$ is clearly visible. 

\section{Nonlinear stability}
\label{sec:nonlinear_stability}

To study the nonlinear stability of global AdS$_4$ with multitrace deformations, we follow Ref.~\cite{Bizon:2011gg} and choose 
\begin{subequations}\label{init_data}
\begin{gather}
\Pi(0,x) = \frac{2\varepsilon}{\pi}\exp\left(-\frac{4 \tan^2(x)}{\pi^2\sigma^2} \right), \\ 
\phi(0,x) = 0, 
\end{gather}
\end{subequations}
as initial data, to then explore the endpoint of their time evolution as the amplitude $\varepsilon$ and width $\sigma$ are varied. We will focus on the case of sharp Gaussian initial data with a width $\sigma$ parametrically smaller than the AdS radius $L=1$, and hence in the following we set $\sigma = 0.1$ unless stated otherwise. Our numerical methods and code convergence tests are presented in App.~\ref{app:numerics}. 

\subsection{Non-relevant deformations}

\subsubsection{Marginal deformations}

\begin{figure}[h!]
\centering
\includegraphics[width=\linewidth]{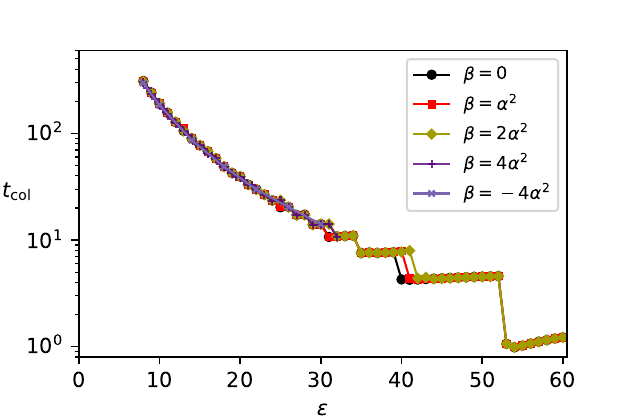}
\caption{Collapse time $t_{col}$ as a function of the initial amplitude $\varepsilon$ for the undeformed case ($\beta = 0$) and four different marginal triple-trace deformations $(\beta\propto\alpha^2)$. The BR instability is not destroyed and the $t_{col}$ behaves as in the undeformed case.}
\label{fig:tcol_marginal}
\end{figure}

Our first main result is that for marginal triple-trace deformations the BR instability is retained, at least as far as our numerical simulations have been able to reach. This fact is illustrated in Fig.~\ref{fig:tcol_marginal}, where we plot the collapse time $t_{col}$ as a function of the initial amplitude $\varepsilon$,  for the case with no deformation ($\beta=0$) and four marginal deformations ($\beta=\alpha^2$, $2 \alpha^2$, $4\alpha^2$ and $-4\alpha^2$). As it is manifest from the figure, $t_{col}(\varepsilon)$ behaves exactly in the same way for these two classes of boundary conditions, and it is compatible with the $\varepsilon \to 0$ scaling  $t_{col}(\varepsilon) \sim \varepsilon^{-2}$ characteristic of the BR instability. 

\begin{figure}[h!]
\centering
\includegraphics[width=\linewidth]{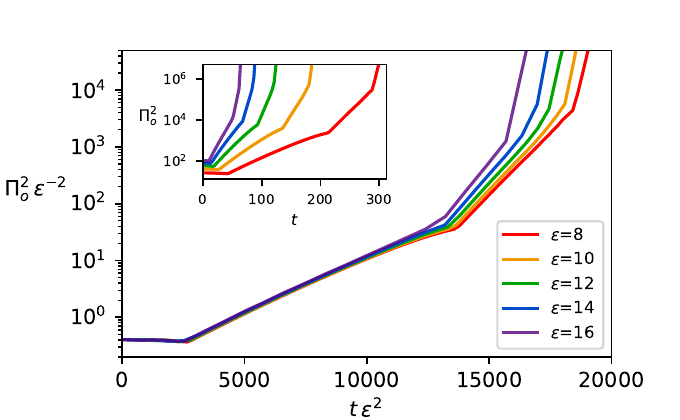}
\caption{For a $\beta = 2\alpha^2$ marginal deformation, the inset shows $\Pi_o^2(t)$ for initial data with progessively smaller $\varepsilon$. The gravitational collapase process is associated with a build-up of this quantity, which acts as a proxy for the Ricci scalar at $x=0$. The main plot shows the rescaled quantities $\varepsilon^{-2} \Pi_o^2$ as functions of the rescaled times $\varepsilon^2 t$. These curves clearly approach a universal form as $\varepsilon$ is reduced, signaling the presence of a BR instability.}
\label{fig:Pio_scaling_marginal}
\end{figure}

Additional evidence for the existence of the BR instability comes from emergent universalities in the bulk dynamics. Following Ref.~\cite{Bizon:2011gg}, in the inset of Fig.~\ref{fig:Pio_scaling_marginal} we show the upper envelope of $\Pi_o^2(t) \equiv \Pi(t, x = 0)^2$ for time evolutions of progressively smaller $\varepsilon$  in the case of a marginal deformation $\beta = 2 \alpha^2$. As it is manifest from the main plot, the rescaled quantities $\varepsilon^{-2}\Pi_o^2(\varepsilon^2 t)$ approach a universal curve as $\varepsilon$ is reduced. In particular, in each case $\varepsilon^{-2} \Pi_o^2(\varepsilon^2 t)$ is upper bounded by its initial value up to a time $\varepsilon^2 t \approx 2630$, after which it starts to grow almost exponentially, to then undergo several stages of progressively amplified growth. The persistence of this behavior in the $\varepsilon \to 0$ limit is tantamount to the nonlinear instability of global AdS$_4$ with a marginal deformation. 

\subsubsection{Irrelevant deformations}

\begin{figure}[h!]
\centering
\includegraphics[width=\linewidth]{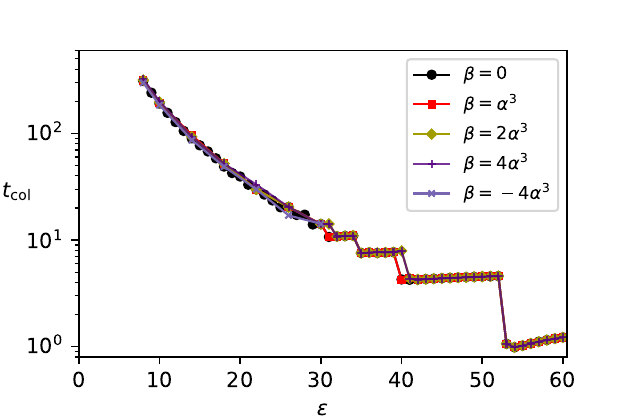}
\caption{Collapse time $t_{col}$ as a function of the initial amplitude $\varepsilon$ for the undeformed case ($\beta = 0$) and four different irrelevant deformations $(\beta\propto\alpha^3)$. The BR instability is not destroyed and collapse time behaves as in the undeformed case.}
\label{fig:tcol_irrelevant}
\end{figure}

For irrelevant deformations of the form $\beta = c_4 \alpha^3$, our numerical results are still compatible with the presence of the BR instability. In Fig.~\ref{fig:tcol_irrelevant}, we plot the collapse time curves for four irrelevant deformations with $c_4 = 1$, $2$, $4$ and $-4$, and compare them with the undeformed $\beta = 0$ case. The five curves are in excellent agreement. 

\begin{figure}[h!]
\centering
\includegraphics[width=\linewidth]{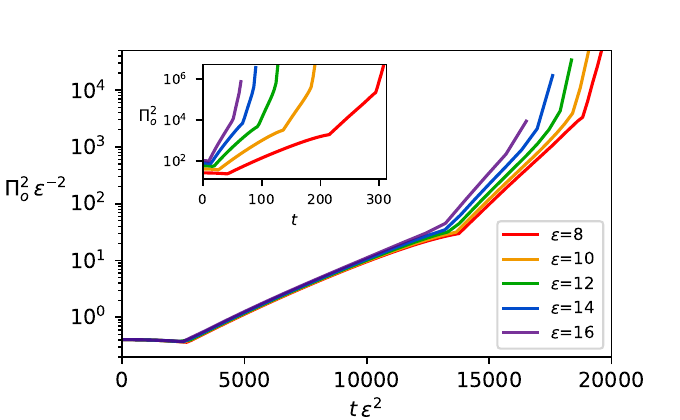}
\caption{For an irrelevant deformation $\beta = 2 \alpha^3$, the inset shows $\Pi_o^2(t)$ for initial data with progressively smaller $\varepsilon$. The main plot shows the rescaled quantities $\varepsilon^{-2} \Pi_o^2$ as functions of the rescaled times $\varepsilon^2 t$. These rescaled quantities clearly approach a universal form as $\varepsilon$ is reduced, signaling the presence of a BR instability.
}
\label{fig:Pio_scaling_irrelevant}
\end{figure}

Additional evidence in favor of the presence of a BR instability comes from a scaling analysis. In Fig.~\ref{fig:Pio_scaling_irrelevant}, we repeat the exercise performed in Fig.~\ref{fig:Pio_scaling_marginal} for an irrelevant deformation $\beta = 2 \alpha^3$. Just as it happened in the previous case, the functions $\varepsilon^{-2}\Pi_o^2(\varepsilon^2 t)$ clearly approach a universal curve as $\varepsilon$ is reduced. 

\subsection{Relevant deformations}

\begin{figure}[h!]
\centering
\includegraphics[width=\linewidth]{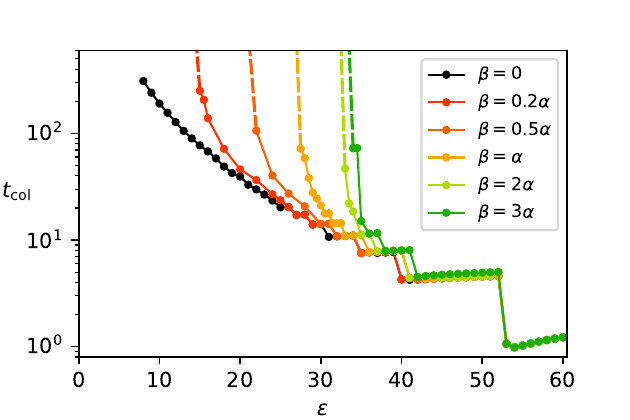}
\includegraphics[width=\linewidth]{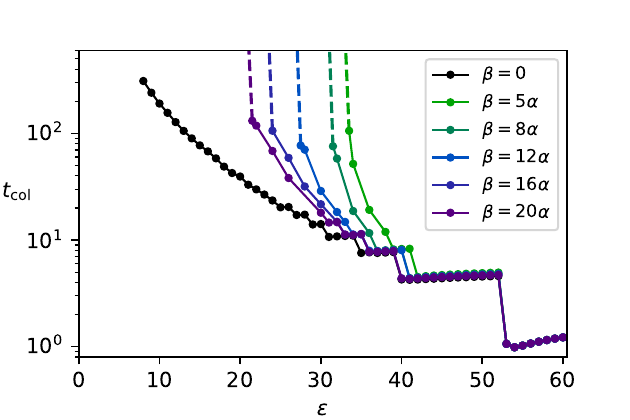}
\caption{Collapse time $t_{col}$ as a function of the initial amplitude $\varepsilon$ for different relevant double-trace deformations, $\beta = c_2\alpha$,  with $c_2 > 0$. The BR instability has disappeared (dashed lines connect to evolutions which have yet not undergone gravitational collapse). \textbf{Top panel.} Upon increasing $c_2$ away from zero, the collapse time curves depart from the $\beta = 0$ one at progressively larger values of $\varepsilon$. \textbf{Bottom panel.} Increasing $c_2$ further, after a critical value the collapse time curves start to lie down again. The reason for this behavior is that, in the $c_2 \to \infty$ limit, we should recover the alternative quantization result, where the BR instability is again present.}
\label{fig:relevant_+}
\end{figure}

\begin{figure}[h!]
\centering
\includegraphics[width=\linewidth]{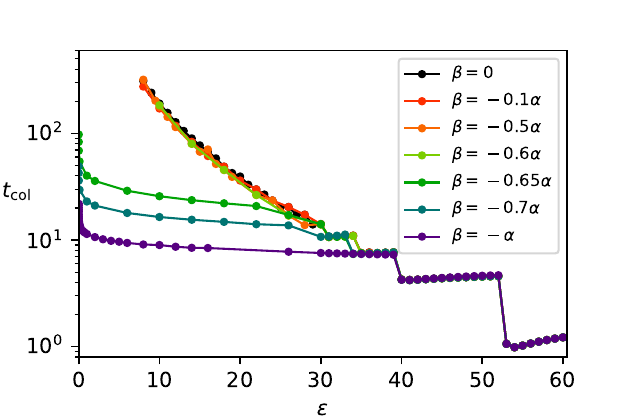}\\
\includegraphics[width=\linewidth]{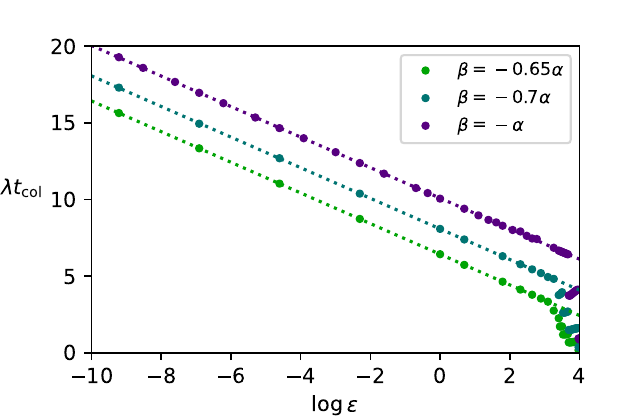}
\caption{\textbf{Top panel.} Collapse time $t_{col}$ as a function of the initial amplitude $\varepsilon$ for different double-trace deformations with $c_2 < 0$. For $-\frac{2}{\pi} < c_2 < 0$, the instability corner is retained as far as our numerical simulations have been able to reach. On the other hand, for $c_2 \leq -\frac{2}{\pi}$, the exponentially growing mode \eqref{unstable_mode} dominates the system dynamics for sufficiently small $\varepsilon$, leading to a qualitatively different collapse time curve. \textbf{Bottom panel.} Demonstration that, for double-trace deformations with $c_2 \leq -\frac{2}{\pi}$, $t_c(\varepsilon) \sim -\log \varepsilon$ in the $\varepsilon \to 0$ limit.}
\label{fig:relevant_-}
\end{figure}

For double-trace deformations, $\beta=c_2 \alpha$, the nonlinear stability properties are markedly different from the non-relevant cases discussed so far. In Fig.~\ref{fig:relevant_+}, we plot the collapse time curves for various double-trace deformations with $c_2 > 0$. In this case, the BR instability is destroyed for any finite value of $c_2$. In the top panel, we observe that the system becomes nonlinearly stable at higher values of $\varepsilon$ as $c_2$ increases. This behavior has a turning point at a critical value of $c_2 \sim 3$. Past this point, and as shown in the bottom panel, the collapse time curves start to lie down again approaching the case without deformation and with a BR instability. This behavior can be explained by noticing that, in the $c_2 \to \infty$ limit, $\frac{\alpha}{\beta} \to 0$ and the double-trace deformation approaches the standard quantization \eqref{stand_quant}, where $\alpha=0$ and the BR instability is at work. 
\squeezeup
% I am here
With $c_2 < 0$, our results for double-trace deformations  are depicted in Fig.~\ref{fig:relevant_-}. In this case, there are two qualitatively different regimes to consider. 

The first regime corresponds to $c_2 < - \frac{2}{\pi}$, and is characterized by the fact that the fundamental mode becomes linearly unstable (cf. Eq.~\eqref{unstable_mode}). This exponentially growing mode governs the system dynamics in the $\varepsilon  \to 0$ limit and overshadows the BR instability, leading to gravitational collapse on a time scale $O(\log \varepsilon)$, parametrically smaller than the original $O(\varepsilon^{-2})$. See the bottom panel of Fig.~\ref{fig:relevant_-} for representative examples of this behavior. 

The second regime corresponds to $-\frac{2}{\pi}<c_2 <0$, and is characterized by a purely real eigenfrequency spectrum. Remarkably, in this situation the double-trace deformation does not render the system nonlinearly stable in the region of initial amplitudes we have been able to probe, despite the fact that the eigenfrequency spectrum is only asymptotically resonant  (see Sec.~\ref{sec:linear_stability}). The persistence of the nonlinear instability for the initial amplitudes we have explored is manifest from the top panel of Fig.~\ref{fig:relevant_-}, which exhibits the collapse time curves for $\beta = -0.1\alpha, -0.5\alpha$ and $-0.6\alpha$. It is fair to say that the numerical construction of these curves is quite resource demanding. This is due to the fact that a large number of global mesh refinements have been introduced to keep the constraints satisfied and the numerical simulation convergent until the very last moments of the collapse (see App.~\ref{app:numerics} for technical details). 

\begin{figure}[h!]
\centering
\includegraphics[width=\linewidth]{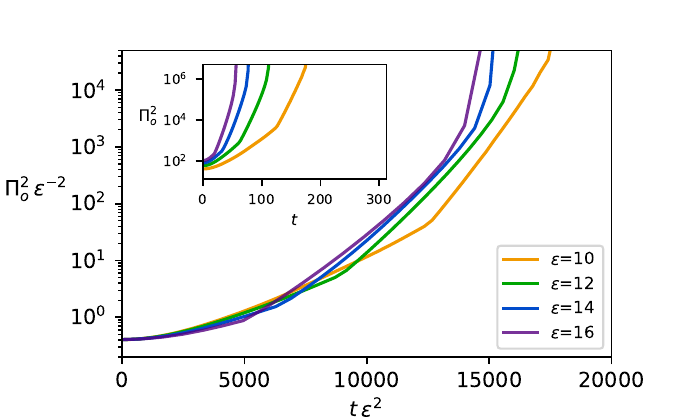}
\caption{For a relevant double-trace deformation $\beta = -0.6 \alpha$, the inset shows $\Pi_o^2(t)$ for initial data with $\varepsilon = 10$, $12$, $14$ and $16$. On the main plot, we represent the rescaled quantities  $\varepsilon^{-2}\Pi_o^2(\varepsilon^2t)$. In contrast to the non-relevant cases depicted in Figs.~\ref{fig:Pio_scaling_marginal} and  \ref{fig:Pio_scaling_irrelevant}, there is no universal behavior emerging when $\varepsilon$ is reduced. This is incompatible with a BR instability being at work in this case.}
\label{fig:Pio_scaling_relevant}
\end{figure}

This unexpected region of nonlinear instability does not correspond to a BR instability, despite what the collapse time curves depicted in Fig.~\ref{fig:relevant_-} might suggest. To illustrate this, in Fig.~\ref{fig:Pio_scaling_relevant} we plot $\varepsilon^{-2} \Pi_o^2(\varepsilon^2 t)$ for $c_2 = -0.6$ and $\varepsilon = 10,~12,~14$ and $16$. In contrast to the cases of non-relevant deformations depicted in Figs.~\ref{fig:Pio_scaling_marginal} and \ref{fig:Pio_scaling_irrelevant}, these rescaled functions do not approach a universal curve as $\varepsilon$ is reduced and, correspondinly, $t_{col}$ does not exhibit an emergent $\varepsilon^{-2}$ scaling in the same limit.\footnote{We note that the results presented in Fig.~\ref{fig:Pio_scaling_relevant} look qualitatively similar to the ones obtained in the flat space case discussed in Ref.~\cite{Maliborski:2012gx} for a non-resonant spectrum (see Fig.~5 in the published version).}  

\begin{figure}[h!]
\centering
\includegraphics[width=\linewidth]{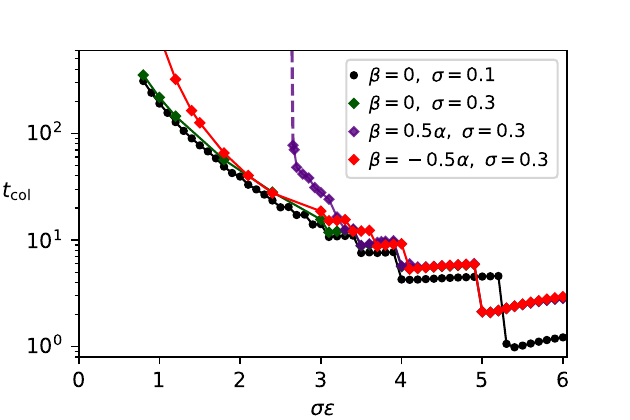}
\caption{Collapse time $t_{col}$ as a function of initial amplitude $\varepsilon$ times initial width $\sigma$, for the undeformed cases with $\sigma=0.1$ and $0.3$, and two relevant deformations of opposite signs, $\beta = 0.5 \alpha$ and $\beta = -0.5 \alpha$, with $\sigma = 0.3$. Both undeformed cases follow the collapse time curve characteristic of the BR instability. In the deformed ones, however, the collapse time curves are clearly lifted.  
}
\label{fig:relevant_wider_-}
\end{figure}

The breakdown of dynamical universality encapsulated by Fig.~\ref{fig:Pio_scaling_relevant} strongly suggests that negative double-trace deformations will render the system nonlinear stable at sufficiently small values of $\varepsilon$, currently sitting outside the reach of our numerical simulations. The results presented in Fig.~\ref{fig:relevant_wider_-} provide further evidence for this hypothesis. In this figure, we explore the effect of double-trace deformations on initial data with a larger width $\sigma=0.3$. With no deformation, the BR instability still exists (for comparison the $\sigma=0.1$ curve has been added). For $c_2 < 0$, however, despite the fact that the system is still nonlinearly unstable in the range of $\varepsilon$ we have been able to probe, the collapse time now deviates significantly from its undeformed counterpart as $\varepsilon$ is reduced. This provides another reason to suspect that a similar deviation will also happen for $\sigma = 0.1$ at sufficiently small values of $\varepsilon$. 

Leaving aside the question of whether the system is nonlinearly stable or not in the $\varepsilon \to 0$ limit, the fact that at finite $\varepsilon$ its behavior is markedly different depending on the sign of $c_2$---as can be seen from Fig.~\ref{fig:relevant_wider_-} and the top panels of Figs.~\ref{fig:relevant_+} and \ref{fig:relevant_-}---is a robust finding of our analysis meriting further study in the search for a definitive explanation. 

\begin{figure}[h!]
\centering
\includegraphics[width=\linewidth]{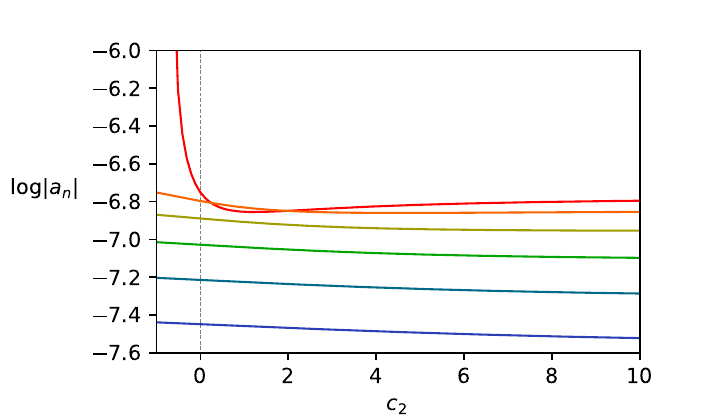}
\caption{First six coefficients $a_1, \ldots, a_6$ of the spectral decomposition \eqref{phi_spectral_decomposition} for the Gaussian initial data \eqref{init_data} ($\varepsilon=1$, $\sigma=0.1$) as functions of the relevant coupling $c_2$. 
}
\label{fig:spectral_decomposition}
\end{figure}

We believe that the ultimate explanation of this difference will involve a detailed understanding of the interplay between the Gaussian initial data \eqref{init_data} and the space of exactly periodic solutions as the relevant coupling $c_2$ is varied. This is motivated by the following observation. Neglecting the backreaction of the scalar field, the spectral decomposition of $\phi(t,x)$ takes the form 
\begin{subequations}\label{phi_spectral_decomposition}
\begin{gather}
\phi(t,x) = \sum_{n=1}^\infty a_n \sin(\omega_n t) e_n(x), \\ 
e_n(x) = \frac{2 \sin(\omega_n x)}{\sqrt{\pi}\sqrt{1-\textrm{sinc}(\pi \omega_n)}\tan(x)}, 
\end{gather}
\end{subequations}
where $\{\omega_n, n \in \mathbb N\}$ are the roots of Eq.~\eqref{eq:eigenfr} and $\{e_n, n\in \mathbb N\}$ a complete set of orthonormal eigenfunctions. Note that, for fixed initial data, the $a_n$ coefficients change as a function of $c_2$, since $\omega_n$, and hence $e_n$, do. The lowest coefficients of the spectral decomposition of the Gaussian initial data \eqref{init_data} (with $\varepsilon = 1$ and $\sigma=0.1$) are provided in Fig.~\ref{fig:spectral_decomposition}. We observe that, while for $c_2 \in \left[-\frac{2}{\pi}, 0\right]$ and $c_2 \to \infty$ the spectral decomposition is dominated by the fundamental mode (with a strong dominance as $c_2\to -\frac{2}{\pi}$), there is a window of positive couplings where the first excited mode dominates.\footnote{The equality between $|a_1|$ and $|a_2|$ is attained at $c_2 \approx 0.24$ and $c_2 \approx 1.97$, and the ratio $|a_1|/|a_2|$ only recovers its value in the alternative quantization at $c_2 \approx 5.46$.} This suggests to us that, for a finite energy, at moderate positive values of $c_2$, the Gaussian initial data \eqref{init_data} can fall into the stability island sourced by a multioscillon analogous to the ones constructed in Ref.~\cite{Masachs:2019znp}. Thinking along these lines, the enhanced nonlinear instability we have found for $c_2 < 0$ could simply be a consequence of the lack of any exactly periodic solution dominated by the fundamental mode for the range of energies we have explored.  
\squeezeup

\squeezeup
\section{Discussion}
\label{sec:discussion}

We have investigated the impact of multitrace deformations on the dynamics of a four-dimensional top-down AdS-Einstein-scalar system. We have discussed how these deformations affect the normal mode eigenfrequency spectrum and linear stability of the system, to then deploy extensive numerical simulations to uncover their effect on the nonlinear stability of Gaussian initial data with width $\sigma$ and amplitude $\varepsilon$. 

For sharp initial data with small $\sigma$, our numerical simulations strongly indicate that marginal and irrelevant multitrace deformations retain the BR instability present in the undeformed case. They also establish that, for double-trace deformations with positive coupling, the BR instability is destroyed and the system is nonlinearly stable for sufficiently small $\varepsilon$. In each case, the presence or absence of the BR instability is perfectly correlated with the exactly or asymptotically resonant character of the eigenfrequency spectrum. 

For double-trace deformations with negative coupling, an exponentially growing normal mode appears below a critical threshold, causing initial data of sufficiently small $\varepsilon$ to undergo gravitational collapse in times parametrically shorter than the ones associated with the BR instability. The most subtle regime we encounter appears for negative couplings above the critical threshold. In this window, and even though the eigenfrequency spectrum is not exactly resonant, the sharpest Gaussian initial data we have considered are still nonlinearly unstable in the whole range of $\varepsilon$ we can simulate reliably. Through a scaling analysis, we have provided strong evidence that this region of enhanced nonlinear instability does not actually correspond to a BR instability, in line with the non-resonant nature of the eigenfrequency spectrum. 

Despite the fact that the numerical evidence at our disposal shows the destruction of the BR instability for double-trace deformations, explaining why the nonlinear stability properties of the system at finite $\varepsilon$ are so different depending on the sign of the coupling is a question we left opened. We believe that the ultimate answer to this question will come from a thorough understanding of how the double-trace coupling affects the space of exactly periodic solutions and the stability islands they source. Exploring this issue  provides a natural avenue for future research. 

\squeezeup
\section*{Acknowledgments}

We would like to thank Anxo Biasi for discussions. 

The simulations in this work were made possible through the access granted by the Galician Supercomputing Center (CESGA) to its infrastructure FinisTerrae III funded by the Spanish Ministry of Science and Innovation, the Galician Government and the European Regional Development Fund (ERDF).
This work has received financial support from Xunta de Galicia (Centro singular de investigaci\'on de Galicia accreditation 2019-2022, and grant ED431C-2021/14), the European Union ERDF, the ``Mar\'\i a de Maeztu" Units of Excellence program MDM-2016-0692, the Spanish Research State Agency (grant PID2020-114157GB-100), and the European Research Council (ERC) under the European Union’s Horizon 2020 research and innovation programme (grant number: 101089093 / project acronym: High-TheQ). Views and opinions expressed are however those of the authors only and do not necessarily reflect those of the European Union or the European Research Council. Neither the European Union nor the granting authority can be held responsible for them. 
The work of DTM has been funded by Xunta de Galicia through the ``Programa de axudas \'a etapa predoutoral da Xunta de Galicia" under the grant ED481A-2020/106, by MICIN through the European Union NextGenerationEU recovery plan (PRTR-C17.I1), and by the Galician Regional Government through the “Planes Complementarios de I+D+I con las Comunidades Aut\'onomas” in Quantum Communication.

\bibliographystyle{bibstyl}
\bibliography{references.bib} % The file containing the bibliography

\providecommand{\href}[2]{#2} \providecommand{\beforedoihref}{} \providecommand{\afterdoihref}{}\begingroup\raggedright\begin{thebibliography}{10}

\bibitem{Bizon:2011gg}
P.~Bizon and A.~Rostworowski, {\it {On weakly turbulent instability of anti-de Sitter space}},  \beforedoihref\href{http://dx.doi.org/10.1103/PhysRevLett.107.031102}{Phys. Rev. Lett.}\afterdoihref\  {\bf 107} (2011) 031102 [\href{http://arXiv.org/abs/1104.3702}{{arXiv:1104.3702}}].

\bibitem{Dias:2012tq}
O.~J.~C. Dias, G.~T. Horowitz, D.~Marolf and J.~E. Santos, {\it {On the Nonlinear Stability of Asymptotically Anti-de Sitter Solutions}},  \beforedoihref\href{http://dx.doi.org/10.1088/0264-9381/29/23/235019}{Class. Quant. Grav.}\afterdoihref\  {\bf 29} (2012) 235019 [\href{http://arXiv.org/abs/1208.5772}{{arXiv:1208.5772}}].

\bibitem{Maliborski2013}
M.~Maliborski and A.~Rostworowski, {\it Time-periodic solutions in an {Einstein} {AdS}{\textendash}massless-scalar-field system},  \beforedoihref\href{http://dx.doi.org/10.1103/PhysRevLett.111.051102}{Phys. Rev. Lett.}\afterdoihref\  {\bf 111} (Aug, 2013) 051102 [\href{http://arXiv.org/abs/1303.3186}{{arXiv:1303.3186}}].

\bibitem{Green:2015dsa}
S.~R. Green, A.~Maillard, L.~Lehner and S.~L. Liebling, {\it {Islands of stability and recurrence times in AdS}},  \beforedoihref\href{http://dx.doi.org/10.1103/PhysRevD.92.084001}{Phys. Rev. D}\afterdoihref\  {\bf 92} (2015), no.~8 084001 [\href{http://arXiv.org/abs/1507.08261}{{arXiv:1507.08261}}].

\bibitem{Buchel:2013uba}
A.~Buchel, S.~L. Liebling and L.~Lehner, {\it {Boson stars in AdS spacetime}},  \beforedoihref\href{http://dx.doi.org/10.1103/PhysRevD.87.123006}{Phys. Rev. D}\afterdoihref\  {\bf 87} (2013), no.~12 123006 [\href{http://arXiv.org/abs/1304.4166}{{arXiv:1304.4166}}].

\bibitem{Maliborski:2014rma}
M.~Maliborski and A.~Rostworowski, {\it {What drives AdS spacetime unstable?}},  \beforedoihref\href{http://dx.doi.org/10.1103/PhysRevD.89.124006}{Phys. Rev. D}\afterdoihref\  {\bf 89} (2014), no.~12 124006 [\href{http://arXiv.org/abs/1403.5434}{{arXiv:1403.5434}}].

\bibitem{Arias:2016aig}
R.~Arias, J.~Mas and A.~Serantes, {\it {Stability of charged global AdS$_{4}$ spacetimes}},  \beforedoihref\href{http://dx.doi.org/10.1007/JHEP09(2016)024}{JHEP}\afterdoihref\  {\bf 09} (2016) 024 [\href{http://arXiv.org/abs/1606.00830}{{arXiv:1606.00830}}].

\bibitem{Masachs:2019znp}
R.~Masachs and B.~Way, {\it {New islands of stability with double-trace deformations}},  \beforedoihref\href{http://dx.doi.org/10.1103/PhysRevD.100.106017}{Phys. Rev. D}\afterdoihref\  {\bf 100} (2019), no.~10 106017 [\href{http://arXiv.org/abs/1908.02296}{{arXiv:1908.02296}}].

\bibitem{Bizon:2020yqs}
P.~Bizo\'n, D.~Hunik-Kostyra and M.~Maliborski, {\it {AdS Robin solitons and their stability}},  \beforedoihref\href{http://dx.doi.org/10.1088/1361-6382/ab7ee4}{Class. Quant. Grav.}\afterdoihref\  {\bf 37} (2020), no.~10 105010 [\href{http://arXiv.org/abs/2001.03980}{{arXiv:2001.03980}}].

\bibitem{Ficek:2023qni}
F.~Ficek and M.~Maliborski, {\it {Dynamics of nonlinear scalar field with Robin boundary condition on the Schwarzschild\textendash{}anti\textendash{}de~Sitter background}},  \beforedoihref\href{http://dx.doi.org/10.1103/PhysRevD.109.044015}{Phys. Rev. D}\afterdoihref\  {\bf 109} (2024), no.~4 044015 [\href{http://arXiv.org/abs/2312.02760}{{arXiv:2312.02760}}].

\bibitem{Maliborski:2012gx}
M.~Maliborski, {\it {Instability of Flat Space Enclosed in a Cavity}},  \beforedoihref\href{http://dx.doi.org/10.1103/PhysRevLett.109.221101}{Phys. Rev. Lett.}\afterdoihref\  {\bf 109} (2012) 221101 [\href{http://arXiv.org/abs/1208.2934}{{arXiv:1208.2934}}].

\bibitem{Kurzweil:2022red}
J.~Kurzweil and M.~Maliborski, {\it {Resonant dynamics and the instability of the box Minkowski model}},  \beforedoihref\href{http://dx.doi.org/10.1103/PhysRevD.106.124020}{Phys. Rev. D}\afterdoihref\  {\bf 106} (2022), no.~12 124020 [\href{http://arXiv.org/abs/2209.05608}{{arXiv:2209.05608}}].

\bibitem{Witten:2001ua}
E.~Witten, {\it {Multitrace operators, boundary conditions, and AdS / CFT correspondence}},  \href{http://arXiv.org/abs/hep-th/0112258}{{arXiv:hep-th/0112258}}.

\bibitem{Papadimitriou:2007sj}
I.~Papadimitriou, {\it {Multi-Trace Deformations in AdS/CFT: Exploring the Vacuum Structure of the Deformed CFT}},  \beforedoihref\href{http://dx.doi.org/10.1088/1126-6708/2007/05/075}{JHEP}\afterdoihref\  {\bf 05} (2007) 075 [\href{http://arXiv.org/abs/hep-th/0703152}{{arXiv:hep-th/0703152}}].

\bibitem{Hertog:2004rz}
T.~Hertog and G.~T. Horowitz, {\it {Towards a big crunch dual}},  \beforedoihref\href{http://dx.doi.org/10.1088/1126-6708/2004/07/073}{JHEP}\afterdoihref\  {\bf 07} (2004) 073 [\href{http://arXiv.org/abs/hep-th/0406134}{{arXiv:hep-th/0406134}}].

\bibitem{Hertog:2005hu}
T.~Hertog and G.~T. Horowitz, {\it {Holographic description of AdS cosmologies}},  \beforedoihref\href{http://dx.doi.org/10.1088/1126-6708/2005/04/005}{JHEP}\afterdoihref\  {\bf 04} (2005) 005 [\href{http://arXiv.org/abs/hep-th/0503071}{{arXiv:hep-th/0503071}}].

\bibitem{Basu:2013soa}
P.~Basu, D.~Das, S.~R. Das and K.~Sengupta, {\it {Quantum Quench and Double Trace Couplings}},  \beforedoihref\href{http://dx.doi.org/10.1007/JHEP12(2013)070}{JHEP}\afterdoihref\  {\bf 12} (2013) 070 [\href{http://arXiv.org/abs/1308.4061}{{arXiv:1308.4061}}].

\bibitem{Ares:2021ntv}
F.~R. Ares, O.~Henriksson, M.~Hindmarsh, C.~Hoyos and N.~Jokela, {\it {Effective actions and bubble nucleation from holography}},  \beforedoihref\href{http://dx.doi.org/10.1103/PhysRevD.105.066020}{Phys. Rev. D}\afterdoihref\  {\bf 105} (2022), no.~6 066020 [\href{http://arXiv.org/abs/2109.13784}{{arXiv:2109.13784}}].

\bibitem{Ares:2021nap}
F.~R. Ares, O.~Henriksson, M.~Hindmarsh, C.~Hoyos and N.~Jokela, {\it {Gravitational Waves at Strong Coupling from an Effective Action}},  \beforedoihref\href{http://dx.doi.org/10.1103/PhysRevLett.128.131101}{Phys. Rev. Lett.}\afterdoihref\  {\bf 128} (2022), no.~13 131101 [\href{http://arXiv.org/abs/2110.14442}{{arXiv:2110.14442}}].

\bibitem{Caddeo:2024lfk}
A.~Caddeo, O.~Henriksson, C.~Hoyos and M.~Sanchez-Garitaonandia, {\it {Spinodal slowing down and scaling in a holographic model}},  \beforedoihref\href{http://dx.doi.org/10.1007/JHEP08(2024)091}{JHEP}\afterdoihref\  {\bf 08} (2024) 091 [\href{http://arXiv.org/abs/2406.15297}{{arXiv:2406.15297}}].

\bibitem{Aharony:2008ug}
O.~Aharony, O.~Bergman, D.~L. Jafferis and J.~Maldacena, {\it {N=6 superconformal Chern-Simons-matter theories, M2-branes and their gravity duals}},  \beforedoihref\href{http://dx.doi.org/10.1088/1126-6708/2008/10/091}{JHEP}\afterdoihref\  {\bf 10} (2008) 091 [\href{http://arXiv.org/abs/0806.1218}{{arXiv:0806.1218}}].

\bibitem{Cvetic:1999xp}
M.~Cvetic, M.~J. Duff, P.~Hoxha, J.~T. Liu, H.~Lu, J.~X. Lu, R.~Martinez-Acosta, C.~N. Pope, H.~Sati and T.~A. Tran, {\it {Embedding AdS black holes in ten-dimensions and eleven-dimensions}},  \beforedoihref\href{http://dx.doi.org/10.1016/S0550-3213(99)00419-8}{Nucl. Phys. B}\afterdoihref\  {\bf 558} (1999) 96--126 [\href{http://arXiv.org/abs/hep-th/9903214}{{arXiv:hep-th/9903214}}].

\bibitem{Costa:2017tug}
M.~S. Costa, L.~Greenspan, J.~Penedones and J.~E. Santos, {\it {Polarised Black Holes in ABJM}},  \beforedoihref\href{http://dx.doi.org/10.1007/JHEP06(2017)024}{JHEP}\afterdoihref\  {\bf 06} (2017) 024 [\href{http://arXiv.org/abs/1702.04353}{{arXiv:1702.04353}}].

\bibitem{Hertog:2004dr}
T.~Hertog and K.~Maeda, {\it {Black holes with scalar hair and asymptotics in N = 8 supergravity}},  \beforedoihref\href{http://dx.doi.org/10.1088/1126-6708/2004/07/051}{JHEP}\afterdoihref\  {\bf 07} (2004) 051 [\href{http://arXiv.org/abs/hep-th/0404261}{{arXiv:hep-th/0404261}}].

\bibitem{Hertog:2004ns}
T.~Hertog and G.~T. Horowitz, {\it {Designer gravity and field theory effective potentials}},  \beforedoihref\href{http://dx.doi.org/10.1103/PhysRevLett.94.221301}{Phys. Rev. Lett.}\afterdoihref\  {\bf 94} (2005) 221301 [\href{http://arXiv.org/abs/hep-th/0412169}{{arXiv:hep-th/0412169}}].

\bibitem{Anabalon:2015xvl}
A.~Anabalon, D.~Astefanesei, D.~Choque and C.~Martinez, {\it {Trace Anomaly and Counterterms in Designer Gravity}},  \beforedoihref\href{http://dx.doi.org/10.1007/JHEP03(2016)117}{JHEP}\afterdoihref\  {\bf 03} (2016) 117 [\href{http://arXiv.org/abs/1511.08759}{{arXiv:1511.08759}}].

\bibitem{Maliborski:2013via}
M.~Maliborski and A.~Rostworowski, {\it {Lecture Notes on Turbulent Instability of Anti-de Sitter Spacetime}},  \beforedoihref\href{http://dx.doi.org/10.1142/S0217751X13400204}{Int. J. Mod. Phys. A}\afterdoihref\  {\bf 28} (2013) 1340020 [\href{http://arXiv.org/abs/1308.1235}{{arXiv:1308.1235}}].

\end{thebibliography}\endgroup

\appendix
\onecolumngrid

\section{Equations of motion and multitrace deformations}
\label{app:eom}

For a time-dependent spherically symmetric solution, we work with the ansatz \eqref{time_ansatz}. The equations of motion resulting from the action \eqref{action} are 
\begin{subequations}
\begin{gather}
\delta' = - \frac{1}{4}\sin x \cos x (\Pi^2 + \Phi^2), \label{eq_delta} \\
f' = (\cot x + 3 \tan x) (1 - f) + 2 \tan x \sinh\left(\frac{1}{2} \phi \right)^2  + f \delta', \label{eq_f} \\
\dot{\Phi} = (f e^{-\delta} \Pi)', \label{Phi_eq} \\
\dot \Pi = \cot^2 x (\tan^2 x  f e^{- \delta} \Phi)' + 2 e^{- \delta} \sec^2 x \sinh(\phi), \label{eq_Pi}
\end{gather}
\end{subequations}
where we have introduced the fields $\Phi = \phi'$ and $\Pi = f^{-1} e^{\delta} \dot{\phi}$. There is an additional equation of motion---the momentum constraint---that holds when the rest do. It is given by 
\begin{equation}
\dot{f} = -\frac{1}{2} \sin x \cos x e^{-\delta} f^2 \Pi \Phi.
\end{equation}
We will utilize this equation to monitor the overall quality of our numerical simulation. 

To implement the boundary conditions corresponding to the multitrace deformation, we start by redefining the fields $\phi$ and $\Phi$ in such a way that their leading-order behavior close to the asymptotic boundary is isolated, 
\begin{subequations}
\begin{gather}
\phi(t,x) = \alpha(t) \cos x + \beta(\alpha(t)) \cos^2 x + \tilde \phi(t,x) \cos^2 x, \label{phi_red} \\
\Phi(t,x) = -\alpha(t) \sin x - 2 \beta(\alpha(t)) \sin x \cos x + \tilde \Phi(t,x) \cos x. \label{Phi_red}
\end{gather}
\end{subequations}
The redefined fields $\tilde{\phi}$ and $\tilde{\Phi}$ satisfy the asymptotic boundary conditions 
\begin{equation}
\tilde \phi\left(t,\frac{\pi}{2}\right) = 0, \quad \tilde \Phi\left(t,\frac{\pi}{2}\right) = 0, 
\end{equation}
which ensure that the original fields $\phi$ and $\Phi$ have the correct asymptotic falloff. The equation of motion for $\tilde \Phi$ is 
\begin{equation}\label{Phitilde_eq}
\dot{\tilde \Phi} = \sec x ( (f e^{-\delta} \Pi)' - \sin x \, \dot \alpha) - 2 \sin x \,  \dot \alpha \frac{d\beta}{d\alpha},
\end{equation}
Note that, after the field redefinition, $\alpha$ becomes a dynamical variable which has to be evolved in time in conjunction with $\Pi$ and $\tilde \Phi$. The equation of motion for $\alpha$ follows from the relation
\begin{equation}\label{alpha_eq}
\dot \alpha = -\Pi'\left(t,\frac{\pi}{2}\right). 
\end{equation}
Taking stock, the final set of equations of motion we work with is given by 
\begin{subequations}
\begin{gather}
\delta' = - \frac{1}{4}\sin x \cos x (\Pi^2 + \Phi^2), \label{seteq_delta}\\
f' = (\cot x + 3 \tan x) (1 - f) + 2 \tan x \sinh\left(\frac{1}{2} \phi \right)^2  + f \delta', \label{seteq_f}\\
\dot{\tilde \Phi} = \sec x ( (f e^{-\delta} \Pi)' + \sin x \, \dot \alpha) + 2 \sin x \,  \dot \alpha \frac{d\beta}{d\alpha}, \label{seteq_Phitilde}\\
\dot \Pi = \cot^2 x (\tan^2 x  f e^{- \delta} \Phi)' + 2 e^{- \delta} \sec^2 x \sinh(\phi), \label{seteq_Pi}\\
\dot \alpha = -\Pi'\left(t,\frac{\pi}{2}\right). \label{seteq_alpha}
\end{gather}
\end{subequations}
In the next section, we explain in detail how we solve this nonlinear PDE system. 

\section{Numerical methods}
\label{app:numerics}

\subsection{Integration strategy}

We start by discussing the integration strategy we employed to solve Eqs.~\eqref{seteq_delta}-\eqref{seteq_alpha}. The initial data consists of the values of $\tilde \Phi(t,x)$, $\Pi(t,x)$ and $\alpha(t)$ at a given time $t$; from these fields, we compute $\Phi(t,x)$ with Eq.~\eqref{Phi_red}, from which $\phi(t,x)$ follows by a spatial integration. With the values of $\phi(t,x)$ and $\Pi(t,x)$ at our disposal, we solve the constraint equations \eqref{seteq_delta}-\eqref{seteq_f} to find $f(t,x)$ and $\delta(t,x)$. We impose the gauge choice $\delta(t,\frac{\pi}{2})=0$, which identifies the time coordinate $t$ with the time on the boundary field theory. Finally, $\tilde \Phi$, $\Pi$ and $\alpha$ are evolved to time $t+ \delta t$ with an explicit fourth-order Runge-Kutta algorithm. Note that the boundary condition $\beta = \beta_{BC}(\alpha)$ is applied implicitly both when obtaining $\Phi(t,x)$ and $\phi(t,x)$ through the redefinition \eqref{Phi_red}, and when using the evolution equation \eqref{seteq_Phitilde} for $\tilde \Phi$. 

\subsection{Technical remarks}

There are critical technical issues that need to be taken into account to obtain a convergent simulation code. 

The first concerns artificial dissipation. Our code employs Kreiss-Oliger (KO) dissipation to damp the numerical noise caused by non-propagating modes with wavelength commensurate with the lattice spacing. The way the KO terms are added in the vicinity of the asymptotic boundary critically affects code convergence. The strategy we followed to obtain a good convergence (see App.~ \ref{app:tests}) was to damp the KO terms by multiplying them with a linear ramp that goes to zero at the asymptotic boundary. Choosing the right spatial extent for the ramp is crucial: when the time evolution demands a global mesh refinement, one needs to keep intact the total number of grid points along which the KO dissipation is damped, rather than keeping the physical length of the ramp fixed---which would double the number of grid points modulated by the ramp at each refinement. 

The second factor influencing the convergence of the simulation code is the method employed to compute  $\Pi'\left(t, \frac{\pi}{2} \right)$ (which is needed to obtain $\dot \alpha$, cf.~Eq.~\eqref{seteq_alpha}). In our implementation, we proceeded as follows. First, from the near-boundary expansion of the equations of motion, we have that (recall that $y = \frac{\pi}{2} - x$)
\begin{equation}\label{asymp_Pi}
\Pi = \dot \alpha y + \dot \beta y^2 + O(y^3). 
\end{equation}
Second, the boundary condition $\beta = \beta_{BC}(\alpha)$ implies that 
\begin{equation}
\dot\beta(\alpha)=\frac{d\beta(\alpha)}{d\alpha}\dot\alpha .
\end{equation}
Combining both observations, we take the ansatz 
\begin{equation}\label{Pi_near-boundary_ansatz}
\Pi(t,x) = \dot \alpha y + \frac{d\beta_{BC}(\alpha)}{d\alpha}\dot\alpha y^2 + a_3 y^3 + a_4 y^4, 
\end{equation}
demand that it agrees with the numerically determined values of $\Pi$ at the grid points $y = n \delta x$ for $n = 1, 2, 3$ (where $\delta x$ is the grid spacing), and solve for $\dot \alpha$, $a_3$ and $a_4$. In the end, we obtain our working expression for $\dot\alpha$,
\begin{equation}
\dot \alpha = -\Pi'(t, \pi/2) = \frac{4\Pi_{\frac{\pi}{2}-3 \delta x} - 27 \Pi_{\frac{\pi}{2}-2 \delta x} +108\Pi_{\frac{\pi}{2}-1 \delta x}}{6 \delta x\left(11+6\frac{d\beta}{d\alpha}\delta x\right)},
\end{equation}
Finally, we note that evaluating the right-hand sides of Eqs.~\eqref{seteq_Phitilde}-\eqref{seteq_Pi} on the discretization grid might produce an indeterminate form $0/0$ leading to numerical instabilities. A useful method to deal with this situation is to apply L'H\^opital's rule, as noted in Ref.~\cite{Maliborski:2013via}. Given two functions $f(x)$ and $g(x)$ such that $f(x)/g(x) \to 0/0$ as $x \to x_0$, the key idea is not to evaluate $f/g$ directly, but rather to replace it by the right-hand side of the identity
\begin{equation}\label{lhopital}
\frac{f}{g} = \frac{f'}{g'} - \frac{g}{g'} \left(\frac{f}{g} \right)'.  
\end{equation}
Contrary to the left-hand side, on the right-hand side the first term is well-behaved by construction and, in the second one, the numerical instabilities induced by the small denominator inside the parenthesis are compensated by the small multiplicative factor in front. 

We apply this procedure to evaluate the right-hand sides of the evolution equation of $\tilde\Phi$ (near the asymptotic boundary) and of $\tilde\Pi$ (both near the origin and near the asymptotic boundary); in the latter case, the method needs to be applied twice. A subtle point here is the way in which the quotient inside the parenthesis in Eq.~\eqref{lhopital} is regularized before taking the derivative. A significant example of this issue arises when we apply this procedure to $\dot{\tilde \Phi}$. The first term on the right-hand side of Eq.~\eqref{seteq_Phitilde},
\begin{equation}\label{term} 
\sec x ( (f e^{-\delta} \Pi)' + \sin x \, \dot \alpha),
\end{equation} 
behaves as $0/0$ for $y \to 0$. This term corresponds to $f/g$ in Eq.~\eqref{lhopital}, and has to be regularized before taking the spatial derivative on the right-hand side of this expression. The value of \eqref{term} at the boundary could be obtained either as a spatial derivative of the field $\Pi$, 
\begin{equation}\nonumber 
-\Pi''\left(t,x=\frac{\pi}{2}\right), 
\end{equation}
or be specified through $\alpha$ and $\beta$ as 
\begin{equation}\nonumber 
-2\frac{d\beta_{BC}(\alpha)}{d\alpha}\dot\alpha, 
\end{equation}
explicitly providing information about the multitrace deformation. We have found that the latter option significantly improves the convergence of the code.

\subsection{Error control}

To assess the quality of the numerical simulations, we monitor the evolution of certain quantities which have to be zero in the continuum limit: the time derivative of the total mass and the integrated momentum constraint. We also qualitatively check that there is no tail of rapid oscillations following the traveling pulse of the scalar field, as this feature typically arises when the discretization near the asymptotic boundary is not optimal (for instance, by not implementing L'H\^opital's rule).    

To keep the numerical simulation in check, we use the integrated momentum constraint to trigger a global mesh refinement when necessary. Simulations undergoing gravitational collapse at very late times with initially small amplitudes are particularly demanding. In these cases, we start with a resolution of $N = 2^{13}+1$ physical grid points and need to go up to resolutions of $N=2^{16}+1$; in fact, the lowest initial amplitude for which we did observe gravitational collapse required a resolution of $N=2^{18}+1$. 

\subsection{Convergence tests}
\label{app:tests}

\begin{figure}[h!]
\centering
\includegraphics[width=0.5\linewidth]{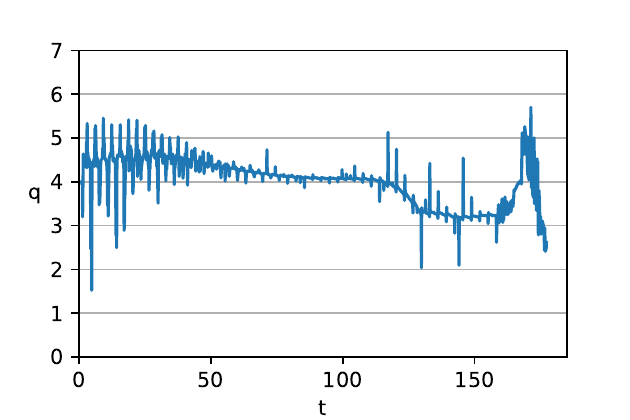}\includegraphics[width=0.5\linewidth]{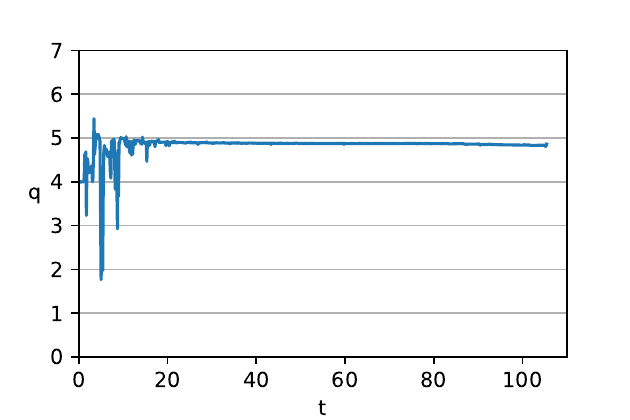}
\includegraphics[width=0.5\linewidth]{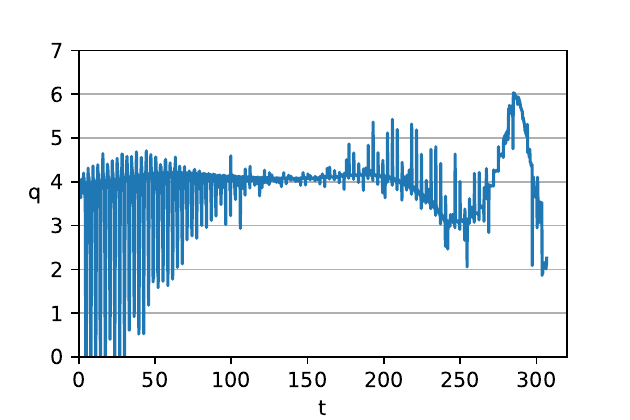}\includegraphics[width=0.5\linewidth]{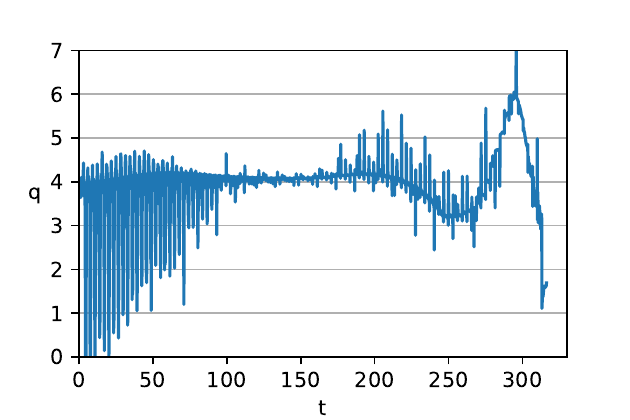}
\caption{\textbf{Top left:} Convergence order $q$ for the evolution of initial data with $\varepsilon=10$, $\sigma=0.1$, and double-trace boundary conditions $\beta =-0.5\alpha$. \textbf{Top right:} Convergence order for initial data with $\varepsilon=33.5$, $\sigma=0.1$ and boundary conditions $\beta=5\alpha$. \textbf{Bottom left:} Convergence order for the evolution of initial data with $\varepsilon=8$, $\sigma=0.1$, and triple-trace (marginal) boundary conditions $\beta =2\alpha^2$. \textbf{Bottom right:} Convergence order for the evolution of initial data with $\varepsilon=8$, $\sigma=0.1$, and irrelevant boundary conditions $\beta =2\alpha^3$.}
\label{fig:convergence}
\end{figure}

We have checked the convergence of our simulation code by evolving the same initial data and boundary conditions using three different grid spacings. Let $g_{a}\left(t,x\right)$ be a certain quantity obtained from the evolution on the grid with spacing $\delta x = \frac{\pi}{2a}$, and consider the norm 
\begin{equation}
\left\Vert g_{a}-g_{b}\right\Vert _{c} \equiv \left(\int_{0}^{\pi/2}dx\left(g_{a}-g_{b}\right)^{2}\right)^{1/2},
\end{equation}
where the sub-index $c$ denotes that the integration is performed on a grid with $\frac{\pi}{2c}$ spacing. We take the convergence rate to be  
\begin{equation}
2^{-q} = \frac{\left\Vert \Pi_{4n}-\Pi_{2n}\right\Vert _{n} + \left\Vert \tilde \Phi_{4n}-\tilde \Phi_{2n}\right\Vert _{n}}{\left\Vert \Pi_{2n}-\Pi_{n}\right\Vert _{n} + \left\Vert \tilde \Phi_{2n}-\tilde \Phi_{n}\right\Vert _{n}},
\end{equation}
where $q$ is the convergence order. Fig.~\ref{fig:convergence} shows the convergence order $q$ for some of the simulations presented in the main text. 

\end{document}